\documentstyle[preprint,revtex,eqsecnum]{aps}
\tightenlines
\begin{document}
\preprint{McGill/92-33.}
\vskip 6 pt
\preprint{July, 1992.}
\draft
\begin{title}
{\bf
Soft dilepton production in relativistic\\
heavy ion collisions}
\end{title}
\author{K. Haglin~\cite{KH}, C. Gale~\cite{CG} and V.
Emel'yanov~\cite{VE1,VE2}}
\begin{instit}
Physics Dept., McGill University, Montr\'eal, P.Q., H3A 2T8, Canada
\end{instit}
\vskip 0.8 true in
\centerline{\bf Abstract}
\vskip -.25 true in
\begin{abstract}
Production of electron-positron pairs with invariant masses less
than 300 MeV from thermalized hadronic matter
in relativistic heavy ion collisions is calculated using a soft virtual
photon approximation.  The general theoretical framework is reviewed and
extended to include arbitrarily massed and charged reaction partners, which
we apply to pions and quarks.
This result, exact within the soft photon approximation, is compared
with a widely used approximate result which uses an electromagnetic amplitude
limited in validity to momentum transfers less than $4m_{\pi(q)}^{2}$.
This momentum-restricted method works very well for pions, constituent
quarks and medium quarks; whereas, it fails when applied
to current quarks.   A field theory calculation is performed
for the $\pi \pi$ elastic cross section which gives excellent
agreement with data.  Quark-antiquark annihilation
diagrams in the Born approximation are estimated and compared with
$\pi^{0}$ and $\eta$ Dalitz decay contributions to the $e^{+}e^{-}$
invariant mass spectra.  A comparison is made between the
rate of production of zero total momentum soft dileptons obtained
using resummation techniques in QCD perturbation theory to that which
we calculate using this soft photon approximation.  Then a Bjorken picture
for the evolution is adopted allowing an integration over
the history of the colliding nuclei.  Using initial conditions
likely to be found at RHIC or LHC we
conclude even if Dalitz decays can be identified and subtracted
from the experimental data it will be difficult to distinguish quark
from pion degrees of freedom in the low-mass $e^{+}e^{-}$ spectra.

\end{abstract}

\narrowtext

\section{Introduction}
\label{sec:intro}

Dileptons and photons~\cite{ef76,es78,es80} provide natural tools for
observing thermalized nuclear matter.  Once produced, they hardly interact
and collide infrequently, if at all, with hadrons in a region
of hot and dense matter.
Therefore, photons and lepton pairs are among the best carriers of
information about the earliest stages of the heavy ion collision.  In
principle, they can carry information about the temperature of the
primordial state.  That there are many competing sources of photons
and dileptons both before and after the hot stage of the
collision makes reconstruction of the hottest state very difficult.
The best that one can do is to evaluate the contribution from
each source to ascertain whether or not it is dominant in some
part of the phase space or if it has distinct properties rendering
separation feasible.  We proceed along these lines by considering four
regions in the invariant mass $M$ of the produced dileptons.  For
invariant masses $M$ above the $J/\psi$ peak, the spectrum is
dominated by Drell-Yan production~\cite{es78}.  The spectrum around
$M \simeq 1$ GeV is dominated by $\rho, \omega,$ and $\phi$ decays.
Dileptons with invariant masses in the region 1.5--3.0 GeV seem to be
mostly thermal~\cite{es80}.  Competing sources in this mass region are
$D\bar D$ decays~\cite{as89}, Drell-Yan and pre-equilibrium
emission~\cite{ke90,ma91}.  Recent calculations~\cite{pr91} of dilepton
spectra at invariant mass 1 GeV $< M <$ 3 GeV show that for
presently achieved energy densities (CERN, BNL) the background
is larger than thermal contributions.  This is somewhat contrary to
earlier calculations~\cite{rh85,kk86} and the reason, in our opinion,
is that the earlier calculations used initial temperatures and energy
densities that were too high.

The last region of invariant mass ($M < m_{\rho}$), which
we will consider in this paper, is a region of {\em soft}
dilepton production~\cite{es80}.  There are many sources
of dileptons with masses $M < m_{\rho}$:  $\pi^{0}$ and
$\eta$ Dalitz decays, $\pi^{+}\pi^{-}$ annihilation
(for $M^{e^{+}e^{-}} > 2m_{\pi}$), pion scattering with
virtual bremsstrahlung and if a quark-gluon plasma (QGP) is formed
there will be quark-antiquark annihilation and quark-quark(antiquark)
scattering with virtual bremsstrahlung.  Recent
calculations~\cite{jc91,kh92}
have shown that in the mass region $M \simeq 2m_{\pi}$ the contribution
from $\eta$ Dalitz decay is two or thee orders of magnitude above
$q \bar q$ annihilation spectra.  But these calculations
of the $q \bar q$ annihilation were done in the Born approximation.  However,
it has been shown~\cite{ebrpty90} in the case of zero $p_{\bot}$ dileptons
that for small mass pairs the perturbative corrections can be several order
of magnitude larger than the Born term.  So, we consider this $q \bar q$
annihilation spectra as a lower limit only.

The contribution of virtual bremsstrahlung from pion scattering
is the same order as $\eta$ Dalitz decay in the mass
region $M \simeq 2m_{\pi}$.  This conclusion was reached by including
only the processes $\pi^{\pm} \pi^{0} \rightarrow \pi^{\pm}
\pi^{0}\gamma^{*}$~\cite{jc91}.   There are other pion-pion reactions, such
as $\pi^{\pm} \pi^{\mp} \rightarrow \pi^{\pm} \pi^{\mp}\gamma^{*}$,
which will contribute to the overall $e^{+}e^{-}$ spectrum.  We will
discuss such processes in this paper.  Typically, a momentum
transfer restriction ($|t| < 4m^{2}$) is the price to pay for a simple but
approximate expression describing the electromagnetic factor in the cross
section.  Such expressions can be quite accurate when the masses are large,
but for small massed particles such as current quarks, the approximation
may not be so good.  In this paper we relax this restriction on $|t|$
by considering scattering at any angles or energies for the charged particles.
We will investigate the influence of such relaxation on the
$e^{+}e^{-}$ production rates and total yields.

If we consider process of $q \bar q$ annihilation in a QGP, it is also
necessary to include quark (antiquark) and/or gluon scattering
with virtual bremsstrahlung and their contribution to pair production.
Their consideration is not possible without knowledge of
the differential cross sections for quark-quark and quark-gluon
elastic scattering.  Using lattice gauge theory data on
screening, we shall estimate the current ($m_{q} \simeq 5$ MeV),
medium ($m_{q} = [2\pi \alpha_{s} /3]^{1/2}T$) and constituent
($m_{q} \simeq 300$ MeV) quark contribution to the electron-positron
mass spectrum.

The paper is organized in the following way.  In Sect.~\ref{sec:picture}
we present the picture of evolution for the colliding nuclear
system.  In Sect.~\ref{sec:softphoton} we discuss the soft photon
approximation for general {\em quasi}-elastic scattering with virtual
bremsstrahlung and calculate the rate of $e^{+}e^{-}$ production.
The two contributing elements are the amplitude for
elastic scattering and the polarization dotted into the electromagnetic
current.  We derive
the exact electromagnetic factor and compare it with a popular
approximate result, limited in validity to $|t| < 4m^{2}$.
In Sect.~\ref{sec:pionbrems} we apply this formalism to thermalized
pions at temperature $T$.  Again, the exact and approximate results
are presented.  Then in Sect.~\ref{sec:quarkbrems} we use our
formulae to calculate pair production through virtual bremsstrahlung in
quark (antiquark)-gluon and quark-quark (antiquark) scattering.
A comparison is made in Sect.~\ref{sec:zeromomentum} among rates
of zero (total) momentum pairs calculated using this soft photon
approximation, QCD perturbation theory and quark-antiquark annihilation
with a modified dispersion relation.
In order to obtain estimates for the total yields from these competing pion
and quark processes, we integrate the rates over the time or
temperature evolution of the colliding nuclei within a Bjorken
picture~\cite{jb83} and present the results in Sect.~\ref{sec:evolut}.
Also in Sect.~\ref{sec:evolut} we use isentropic evolution approximations
to estimate $\pi^{0}, \eta$ Dalitz decays and $q \bar q$ annihilation
in the Born approximation.  Finally, in Sect.~\ref{sec:concl}
we write a brief summary and present concluding remarks.

\section{Evolution of the Colliding Nuclear System}
\label{sec:picture}

We assume formation of thermalized hadronic matter in relativistic
heavy ion collisions.  This approach has recently been popularized
especially with respect to QGP formation.  There is
evidence that in such collisions complete equilibration occurs within
1--2 fm/c~\cite{kg92}.  The idea
of thermalization is attractive because it enables us to use
powerful methods from equilibrium thermodynamics.
If the initial temperature $T_{i} > T_{c} \simeq 200$
MeV, then we expect hadrons to dissolve into freely propagating quarks,
antiquarks and gluons after having gone through a deconfinement phase
transition.  It is well known~\cite{es80} that chiral invariance is
restored near the phase transition temperature, whereupon constituent
quarks ($m_{q} \simeq 300$ MeV) become current quarks with masses
$m_{q} \simeq 5$ MeV.  On the other
hand, in a QGP the fermion mass induced by many-body effects is~\cite{es80}
$m_{q} = [2\pi \alpha_{s}(T)/3]^{1/2}T$.  If we adopt the lattice gauge
theory renormalization group result~\cite{fk88}
\begin{equation}
\alpha_{s} (T) = {6 \pi \over (33-2N_{f})\ln(aT/T_{c})},
\end{equation}
with $a \sim 8$, then for relevant temperatures $T \simeq$
200--300 MeV the medium quark masses $m_{q} \simeq$ 100--150 MeV.
So, if we want to calculate the bremsstrahlung contribution from
the quark phase, it is necessary to take into account the
processes of current and medium quark scattering at
temperatures $T > T_{c}$ as well as constituent quark scattering at
$T=T_{c}$.  During a time evolution the nuclear matter cools.  If
the QGP-hadron phase transition is first order at $T=T_{c}$, the
mixed phase may be realized.  In this case there is dilepton
emission from current, medium, and constituent quarks, as well as from
pion scattering with virtual bremsstrahlung.  At $T=T_{c}$
nuclear matter transforms into an hadronic state, mainly of pions.
For temperatures $T_{c} > T > T_{f} \simeq 140$ MeV, pion
scattering with virtual quanta emission contributes.  At the break-up
temperature $T_{f}$ we are left with only freely streaming particles.

\section{Soft virtual photon approximation}
\label{sec:softphoton}

To begin the discussion, let us first recall some facts about
bremsstrahlung in hadron-hadron collisions.  If charged particles
undergo acceleration, photons (real or virtual) may be radiated.
Soft photon emission occurs only from the external lines~\cite{rr76},
as depicted in Fig~\ref{generice+e-}.
Relative to this, radiation from the strongly interacting blob is
negligible.  It continues to be negligible as long as the energy carried
by the photon is less than the inverse of the strong interaction collision
time $\tau \simeq$ 1--2 fm/c; that is, if $E_{\gamma} \leq$ 100--200
MeV.  The advantage of the soft photon approximation is that the
electromagnetic and strong interaction components
of the matrix elements disentangle leaving separate multiplicative
factors.  This allows
the invariant cross section for scattering and
at the same time producing a soft real photon of four momentum
$q^{\mu}$ to be written as~\cite{rr76}
\begin{equation}
q_{0}{d^{4} \sigma^{\gamma} \over d^{3}q dx} = {\alpha \over 4\pi^{2}}
\left\lbrace \sum\limits_{\rm pol \atop \lambda} J \cdot \epsilon_{\lambda}
\ J \cdot \epsilon_{\lambda} \right\rbrace {d\sigma \over dx},
\label{eq:d4sd3qdx}
\end{equation}
where $d\sigma/dx$ is the strong interaction cross section for the reaction
$a b \rightarrow c d$, $\epsilon_{\lambda}$ is the polarization of the
emitted photon, and
\begin{equation}
J^{\mu} = -Q_{a} {p_{a}^{\mu} \over p_{a}\cdot q} -
Q_{b} {p_{b}^{\mu} \over p_{b}\cdot q} +
Q_{c} {p_{c}^{\mu} \over p_{c}\cdot q}
+ Q_{d} {p_{d}^{\mu} \over p_{d}\cdot q}
\label{eq:jmu}
\end{equation}
is the current. The $Q$\/s and $p$\/s represent the charges (in units
of the proton charge) and the four-momenta of the particles, respectively.
For soft virtual photons (dileptons) it is possible to continue from
$q^{2}=0$ to $q^{2} = M^{2}$, $M$ being the invariant mass of
the $e^{+}e^{-}$ pair.  The result is~\cite{rr76} ($q^{\mu} =
p_{+}^{\mu} + p_{-}^{\mu}$):
\begin{equation}
E_{+}E_{-} {d^{6}\sigma^{e^{+}e^{-}}\over d^{3}p_{+}d^{3}p_{-}}
= {\alpha \over 2 \pi^{2}} {1 \over q^{2}}q_{0}
{d^{3}\sigma^{\gamma} \over d^{3}q}.
\label{eq:extrapolate}
\end{equation}
Eqs.~(\ref{eq:d4sd3qdx})--(\ref{eq:extrapolate}) can be combined
to give the cross section for quasi-elastic $ab \rightarrow cd$ scattering
while at the same time producing an $e^{+}e^{-}$ pair.
Four momentum conservation is strictly enforced at the photon-dilepton
vertices by writing
\begin{eqnarray}
E_{+}E_{-} {d\sigma^{e^{+}e^{-}}_{ab \rightarrow cd} \over
d^{3}p_{+}d^{3}p_{-}} &=&
{\alpha^{2} \over 8\pi^{4}} {1 \over M^{2}} \int
\left|\epsilon \cdot J \right|^{2}_{ab \rightarrow cd}
{d\sigma_{ab \rightarrow cd}\over dt}
\delta^{4}\left(q-(p_{+}+p_{-})\right)d^{4}q dt,
\label{eq:pairprod}
\end{eqnarray}
where $t$ is the four-momentum transfer in the $ab \rightarrow cd$
collision.  We note that this is somewhat different from the expression
used by Gale and Kapusta~\cite{cg87} but this requires no further
approximations whereas in theirs $q_{0}^{2}$ was replaced with
$(E_{+}+E_{-})\sqrt{(E_{+}+E_{-})^{2}-M^{2}}$ in order to facilitate
analytic integration.  Such a replacement introduces a slightly
different mass dependence into the result than does Eq.~\ref{eq:pairprod}.
The expression from Ref.~\cite{cg87} is larger than Eq.~\ref{eq:pairprod}
for small invariant masses and smaller for large invariant masses.  A
region of intermediate mass exists where the two are nearly the same.

Separation of the full amplitude into strong and electromagnetic
multiplicative factors is convenient.  The strong interaction on-shell
cross sections can be parametrized or modeled within, say, a quantum field
theory.  The electromagnetic factor can also be calculated exactly.
It represents a coherent sum of all radiation fields involved, both
initial and final charged states.  Consider the completely general
scattering process
\begin{equation}
ab \rightarrow cd
\end{equation}
where the masses $m_{a}=m_{c}$ and $m_{b}=m_{d}$ are arbitrary
and the charges are subject only to overall charge
conservation, namely, $Q_{a}+Q_{b}=Q_{c}+Q_{d}$.
We calculate the exact squared modulus of the polarization dotted
into the current to be
\begin{eqnarray}
|\epsilon \cdot J|^{2}_{ab\rightarrow cd}
&=& \left. {1\over q_{0}^{2}} \right\lbrace
-( Q_{a}^{2} + Q_{b}^{2} + Q_{c}^{2} + Q_{d}^{2}) \nonumber\\
&\ &- 2(Q_{a}Q_{b}+Q_{c}Q_{d})
{s-m_{a}^{2}-m_{b}^{2} \over \sqrt{[s-(m_{a}-m_{b})^{2}]
[s-(m_{a}+m_{b})^{2}]}} \nonumber\\
&\ & \ \ \ \ \times
\ln \left(
{\sqrt{s-(m_{a}-m_{b})^{2}} + \sqrt{s-(m_{a}+m_{b})^{2}} \over
\sqrt{s-(m_{a}-m_{b})^{2}} - \sqrt{s-(m_{a}+m_{b})^{2}}} \right)
\nonumber\\
& \ & + 2(Q_{a}Q_{c} + Q_{b}Q_{d}) {2m_{a}m_{b}-t \over
\sqrt{t(t-4m_{a}m_{b})}} \nonumber\\
&\ & \ \ \ \ \times
\ln \left( {\sqrt{4m_{a}m_{b}-t} + \sqrt{-t} \over \sqrt{4m_{a}m_{b}-t}
- \sqrt{-t}}
\right)  \nonumber\\
& \ & + 2(Q_{a}Q_{d} + Q_{b}Q_{c}) {s+t-m_{a}^{2}-m_{b}^{2}
\over \sqrt{[s+t-(m_{a}+m_{b})^{2}][s+t-(m_{a}-m_{b})^{2}]}} \nonumber\\
&\ & \ \ \ \ \times \left.
\ln \left( {\sqrt{s+t-(m_{a}-m_{b})^{2}} + \sqrt{s+t-(m_{a}+m_{b})^{2}}
\over \sqrt{s+t-(m_{a}-m_{b})^{2}} - \sqrt{s+t-(m_{a}+m_{b})^{2}}}
\right) \right\rbrace.
\label{eq:ejsquared}
\end{eqnarray}
Details of its lengthy derivation can be found in
Appendix~\ref{app:edotjcorr}.  The usefulness of this formula
is in its generality.  It is a benchmark
with which to compare approximate expressions for $|\epsilon \cdot J|^{2}$,
which is our next task.  If in this previous scattering process,
the momentum transfer is small relative to the masses in the
problem, specifically if $|t| < (m_{a}+m_{b})^{2}$, then a good
approximation to the electromagnetic amplitude is
(for details see Appendix \ref{app:edotjapprox})
\begin{equation}
\left| \epsilon \cdot J\right|^{2}_{ab\rightarrow cd}
 \simeq {2 \over 3} {1\over q_{0}^{2}}
\left( {-t\over m_{a}m_{b}}\right) \left[ ( Q_{a}^{2}
+ Q_{b}^{2}) -{3\over 2} Q_{a} Q_{b}f(s) \right],
\label{eq:approxejsquared}
\end{equation}
where
\begin{eqnarray}
f(s) &=& {w_{-}\over 2w_{+}} - {w_{+}\over 2w_{-}}
-{(w_{-}-w_{+})\over 4w_{-}}\left\lbrace {2\sqrt{w_{-}}\over \sqrt{w_{+}}} +
\left({\sqrt{w_{-}}\over \sqrt{w_{+}}}\right)^{3} \right. \nonumber\\
&\ & \ \ \ \ \ \left. +{\sqrt{w_{+}}\over \sqrt{w_{-}}} \right\rbrace
\ln \left( {\sqrt{w_{-}}+\sqrt{w_{+}} \over \sqrt{w_{-}}
-\sqrt{w_{+}}} \right),
\label{eq:fofsab}
\end{eqnarray}
where $w_{-} \equiv s-(m_{a}-m_{b})^{2}$ and $w_{+}\equiv s-(
m_{a}+m_{b})^{2}$.
The interference function is shown in Fig.~\ref{plotfofs} for
equal-mass scattering $m_{a}=m_{b}=m$.
Its behavior is an amusing consequence of multipole interference.
In the limit $\sqrt{s}/m \rightarrow 2$, $f(s) \rightarrow
-2/3$ and as $\sqrt{s}/m \rightarrow \infty$, $f(s) \rightarrow 0$.
Eqs.~(\ref{eq:approxejsquared}) and (\ref{eq:fofsab})
are extremely useful because the result
is linear in $t$.  It turns out that this approximation is quite
good as long as particles are at least as massive as pions.  For
processes involving lighter particles, one must resort to
Eq.~(\ref{eq:ejsquared}).

Whichever expression is used, either
Eq.~(\ref{eq:ejsquared}) or Eq.~(\ref{eq:approxejsquared}),
the common $1/q_{0}^{2}$ dependence allows analytic integration over
the virtual photon and lepton momenta when inserted into
Eq.~(\ref{eq:pairprod}).  The
resulting differential cross section is
\begin{eqnarray}
{d\sigma^{e^{+}e^{-}}_{ab\rightarrow  cd} \over dM^{2}} & = & {\alpha^{2}
\over 4\pi^{2}}
{\widehat \sigma(s) \over M^{2}} \left\lbrace
\ln \left( {\left(\sqrt{s}-(m_{a}+m_{b})\right) +
\sqrt{\left(\sqrt{s}-(m_{a}+m_{b})\right)^{2}-M^{2}} \over
M}\right) +
\right.  \nonumber\\
&\ & \ \ \ \ \left.
\sqrt{1 - {M^{2}\over \left(\sqrt{s}-(m_{a}+m_{b})\right)^{2}}}
\right\rbrace,
\nonumber\\
\label{eq:sigab}
\end{eqnarray}
where we have defined
\begin{equation}
{\widehat \sigma(s)} \equiv
\int\limits^{0}_{-\lambda(s,m_{a}^{2},m_{b}^{2})/s}
{\hskip -.7 true cm} dt {d\sigma_{ab\rightarrow cd} \over dt}
\left( q_{0}^{2}|\epsilon \cdot J|^{2}_{ab\rightarrow cd} \right).
\label{eq:hatsigma}
\end{equation}
Sometimes called the kinematic {\em triangle function},
$\lambda(x,y,z) = x^{2}-2(y+z)x+(y-z)^{2}$~\cite{bk73}.
Evaluation of the $e^{+}e^{-}$ cross section proceeding along these
lines neglects the momentum of the virtual photon in the phase-space
$\delta$-function for the on-shell elastic cross section.  Strictly
speaking, one should correct for this by including the factor
\begin{equation}
{R_{2}(s_{2},m_{a}^{2},m_{b}^{2}) \over R_{2}(s,m_{a}^{2},m_{b}^{2})} =
{\lambda^{1/2}(s_{2},m_{a}^{2},m_{b}^{2}) \over
\lambda^{1/2}(s,m_{a}^{2},m_{b}^{2})}
{s \over s_{2}}
\label{eq:phasespace}
\end{equation}
(with $s_{2} = s + M^{2} - 2\sqrt{s}q_{0}$) before integrating
Eq.~(\ref{eq:pairprod})~\cite{cgjk89}.  We will include this correction
in the numerical part of this study.
The independent particle approximation from kinetic theory allows
the rate (number of reactions per unit time per unit volume) to
be written as
\begin{eqnarray}
{dN^{e^{+}e^{-}}_{ab}\over d^{4}x dM^{2}} &=& g_{ab}
\int ds\int {d^{3}p_{a}\over (2\pi)^{3}}\int {d^{3}p_{b}\over (2\pi)^{3}}
e^{-\beta(E_{a}+E_{b})} {d\sigma_{ab}^{e^{+}e^{-}}\over dM^{2}}
v_{rel} \delta\left(s-(p_{a}+p_{b})^{2}
\right),
\nonumber\\
\label{eq:epemprodrate}
\end{eqnarray}
where
\begin{eqnarray}
v_{rel} &=& {\sqrt{(p_{a}\cdot p_{b})^{2} - m_{a}^{2}m_{b}^{2}} \over
E_{a}E_{b}},
\nonumber
\end{eqnarray}
$\beta$ is the inverse temperature and $g_{ab}=(2s_{a}+1)(2s_{b}+1)
N_{c}^{a}N_{c}^{b}$ is the spin and color degeneracy.
Integrating over both elastic scattering particles' momenta (for calculational
details see Appendix~{\ref{app:dne+e-d4xdm2}) yields
\begin{eqnarray}
{dN^{e^{+}e^{-}}_{ab}\over d^{4}x dM^{2}} &=& {T^{6}g_{ab}\over
16\pi^{4}} \int\limits_{z_{min}}^{\infty} dz
{\lambda(z^{2}T^{2},m_{a}^{2},m_{b}^{2}) \over T^{4}}
{\cal K}_{1}(z) {d\sigma_{ab}^{e^{+}e^{-}} \over
dM^{2}}(z)
\label{eq:genrate}
\end{eqnarray}
where $z_{min} = (m_{a} + m_{b} + M)/T$, $z=\sqrt{s}/T$, and ${\cal K}_{1}$
is a modified Bessel function.
We are now in position to calculate the rate of electron-positron
production through virtual bremsstrahlung from pion processes,
quark-quark or quark-gluon processes, etc.

\section{Pair production from thermalized pions}
\label{sec:pionbrems}

A truncated sum of the possible pion processes is used,
that is,
\begin{eqnarray}
\sum\limits_{ab, cd}
|\epsilon \cdot J|^{2}_{ab\rightarrow cd} \left({d\sigma
\over dt}\right)_{ab\rightarrow cd} &=&
|\epsilon \cdot J|^{2}_{\pi^{+}\pi^{0}} \left({d\sigma \over dt}
\right)_{\pi^{+}\pi^{0}} \nonumber\\
&\ & +|\epsilon \cdot J|^{2}_{\pi^{-}\pi^{0}}
\left({d\sigma \over dt} \right)_{\pi^{-}\pi^{0}} +
|\epsilon \cdot J|^{2}_{\pi^{+}\pi^{-}}
\left({d\sigma \over dt} \right)_{\pi^{+}\pi^{-}}.
\label{eq:pionsum}
\end{eqnarray}
We refer to this as truncated because with it we are
neglecting $\pi^{+}\pi^{+}$ and $\pi^{-}\pi^{-}$ processes.
Such an omission is justified by the fact that there is significant
cancellation due to destructive interference in like-charge
processes~\cite{klh91}.
We have also neglected $\pi^{+}\pi^{-}\rightarrow
\pi^{0}\pi^{0}$, since $\sigma(\pi^{+}\pi^{-}\rightarrow \pi^{0}\pi^{0})$
is rather small relative to the other pion processes~\cite{jfg76}.
Using the approximate electromagnetic amplitude, Eq.~(\ref{eq:pairprod})
becomes
\begin{eqnarray}
{d\sigma^{e^{+}e^{-}}_{\pi \pi} \over dM^{2}} & = & {\alpha^{2}
\over 6\pi^{2}}
{\bar \sigma_{\pi\pi} (s) \over M^{2}} \Gamma(\sqrt{s},m_{\pi},M)
\left\lbrace {4 \over 3} +
{1 \over 2}f(s) \right\rbrace,
\label{eq:sigpipi}
\end{eqnarray}
where $\sigma_{\pi \pi} \sim \sigma_{\pi^{+}\pi^{0}} +
\sigma_{\pi^{-}\pi^{0}} + \sigma_{\pi^{+}\pi^{-}}$,
the momentum transfer weighted cross section is $\bar \sigma (s)$,
and we have defined yet another function
\begin{eqnarray}
\Gamma(\sqrt{s},m_{\pi},M) &=&
\ln \left( {\left(\sqrt{s}-2m_{\pi}\right) +
\sqrt{\left(\sqrt{s}-2m_{\pi}\right)^{2}-M^{2}} \over
M}\right) +  \nonumber\\
&\ & \ \ \ \ \
\sqrt{1 - {M^{2}\over \left(\sqrt{s}-2m_{\pi}\right)^{2}}}.
\end{eqnarray}
If $d\sigma/dt$ is a symmetric function of $t$ and $u$, then~\cite{jc91}
\begin{equation}
\bar \sigma(s) = 2\sigma_{el}(s) \left[{s\over 4m_{\pi}^{2}} - 1 \right].
\label{eq:barsig}
\end{equation}
The elastic pion-pion cross section is parametrized in the following
way:~\cite{jc91}

a) for $\sqrt{s} \leq$ 0.6 GeV the chiral model expression is used,
\begin{equation}
\sigma_{el}(s) = {2\over 3} {1\over F_{\pi}^{4}} {1\over
16\pi }s \left[ 1 - {5m_{\pi}^{2}\over s} + {7m_{\pi}^{4}\over s^{2}}
\right]
\end{equation}
with the pion decay constant $F_{\pi}= 0.098$ GeV.

b) at the collision energy near the rho mass,  $0.6< \sqrt{s} \leq$
1.5 GeV, the largest contribution to $\pi \pi$ the scattering
amplitude is due to resonance formation.  Therefore,
\begin{equation}
\sigma_{el}(s) = {g_{\rho \pi \pi}^{4} \over 48 \pi s}
{ (s-4m_{\pi}^{2})^{2} \over (s-m_{\rho}^{2})^{2}+
m_{\rho}^{2}\Gamma_{\rho}^{2}},
\end{equation}
where the coupling constant $g_{\rho \pi \pi} \simeq 6$,
$m_{\rho} = 0.775$ GeV, and $\Gamma_{\rho}= 0.155$ GeV.

c) for large collision energy
$\sqrt{s} >$ 1.5 GeV, $\sigma_{el}$ becomes energy independent
$\sigma_{el} \simeq 5$ mb.

With a parametrization of the cross section we can
combine Eqs.~(\ref{eq:genrate}) and ~(\ref{eq:sigpipi}) to calculate
the rate of $e^{+}e^{-}$ production from the sum of
$\pi^{+} \pi^{0}, \pi^{-}\pi^{0}$ and $\pi^{+}\pi^{-}$
processes.  Different regions of $\sqrt{s}$
are investigated by varying the upper
limit on the energy.  The results at $T=200$ MeV are presented in
Fig.~\ref{chiral} along with a comparison of the rate using
a quartic pion interaction ${\cal L}_{I} = \lambda (\vec \pi \cdot
\vec \pi )^{2}/4$
($\lambda=1.4$ from $\pi \pi$ scattering lengths)~\cite{aw91}.
Our results reproduce the rate from the quartic interaction
in the low mass region $M < 100$ MeV only if $\sqrt{s}$ is restricted
to the non-resonance region, say $\sqrt{s} \leq 0.48$ GeV.
Electron-positron pairs with invariant masses larger than $\sim 100$ MeV
are attributed to $\sqrt{s}$ in the resonance region and beyond.

Let us compare this production rate to a result which uses
the exact electromagnetic factor. It is given by
Eq.~(\ref{eq:ejsquared}) with $Q_{a}, Q_{b}, Q_{c}, Q_{d} =
0,\pm 1$ and $m_{a}=m_{b}=m_{\pi}$. Since the $t$ dependence is no longer
linear, knowledge of the energy dependent total elastic scattering cross
section is not enough.  We must have a parametrization or a model
calculation  of the differential cross section $d\sigma_{el}/dt$
from which we integrate to obtain $\sigma_{el}(s)$.
We perform a field theory calculation using $\sigma$, $\rho$, and
$f(1270)$ meson-exchange to model the strong interaction.  The quantum
numbers for these particles are ($I=0,J=0$), ($I=1,J=1$), and ($I=0,J=2$)
respectively.  Low-energy $\pi\pi$ scattering length analyses
suggest that momentum-dependent scalar and tensor couplings to the chiral
fields should be used~\cite{jfd89}.
So the effective Lagrangian we use is
\begin{equation}
{\cal L}_{\rm int} = g_{\sigma} \sigma \partial_{\mu} \vec \pi \cdot
\partial^{\mu} \vec \pi + g_{\rho} \vec \rho^{\ \mu} \cdot (\vec \pi
\times \partial_{\mu} \vec \pi) + g_{f} f_{\mu\nu} \partial^{\mu} \vec \pi
\cdot \partial^{\nu} \vec \pi.
\end{equation}
Suppose we wish to calculate $\sigma(\pi^{+}\pi^{-}\rightarrow
\pi^{+}\pi^{-})$.
The differential cross section is proportional to the square of
the matrix element describing the overlap of initial and final
two-hadron states.  There are six terms in this matrix element:
$t$ and $s$ channel $\sigma$-exchange, $\rho$-exchange and $f$-exchange
processes.
The composite nature of the mesons necessitates modification at high
momentum transfers. The net result is an effective suppression
in that regime, so the vertices in the $t$-channel diagrams
are given momentum-transfer damping monopole form factors
\begin{equation}
h_{\alpha}(t) = {m_{\alpha}^{2} - m_{\pi}^{2} \over  m_{\alpha}^{2} - t},
\end{equation}
where $m_{\alpha}$ is either the $\sigma$, $\rho$, or the $f$ mass depending
upon which meson is exchanged.

In this model finite resonance lifetimes are incorporated into the scalar,
vector, and tensor boson propagators.
For the $f$ propagator we use~\cite{hvd70}
\begin{equation}
i {\cal P}^{\mu\nu\alpha\beta} =
{-i\left\lbrace {1\over 2} \left({1\over 3}g^{\mu\nu}g^{\alpha\beta}
-g^{\mu\alpha}g^{\nu\beta} - g^{\mu\beta}g^{\nu\alpha}\right)
\right\rbrace \over k^{2}-m_{f}^{2} + im_{f}\Gamma_{f}}.
\end{equation}
As previously mentioned, there are six terms comprising
the matrix element.
\begin{equation}
{\cal M} = {\cal M}_{1}+{\cal M}_{2}+{\cal M}_{3}+{\cal M}_{4}
+{\cal M}_{5}+{\cal M}_{6},
\end{equation}
where
\begin{eqnarray}
{\cal M}_{1} &=& {-g_{\sigma}^{2} h_{\sigma}^{2}(t)(2m_{\pi}^{2}-t)^{2}
\over t-m_{\sigma}^{2}+im_{\sigma}\Gamma_{\sigma}} \nonumber\\
{\cal M}_{2} &=& {-g_{\sigma}^{2} (s-2m_{\pi}^{2})^{2} \over
s-m_{\sigma}^{2}+im_{\sigma}\Gamma_{\sigma}} \nonumber\\
{\cal M}_{3} &=& {-g_{\rho}^{2} h_{\rho}^{2}(t) (s-u) \over
t-m_{\rho}^{2}+im_{\rho}\Gamma_{\rho}} \nonumber\\
{\cal M}_{4} &=& {g_{\rho}^{2} (u-t) \over
s-m_{\rho}^{2}+im_{\rho}\Gamma_{\rho}}\nonumber\\
{\cal M}_{5} &=& {g_{f}^{2} h_{f}^{2}(t) \over
t-m_{f}^{2}+im_{f}\Gamma_{f}} {1\over 2}\left({1\over 3}
(2m_{\pi}^{2}-t)^{2}-(s-2m_{\pi}^{2})^{2}-(2m_{\pi}^{2}-u)^{2}
\right) \nonumber\\
{\cal M}_{6} &=& {g_{f}^{2} \over
s-m_{f}^{2}+im_{f}\Gamma_{f}} {1\over 2}\left({1\over 3}
(s-2m_{\pi}^{2})^{2}-(2m_{\pi}^{2}-t)^{2}-(2m_{\pi}^{2}-u)^{2}
\right). \nonumber\\
\end{eqnarray}
Notice that as written $g_{\sigma}$ and $g_{f}$ are not dimensionless but
$g_{\sigma}m_{\sigma}$ and $g_{f}m_{f}$ are.
The full amplitude is symmetric under the interchange of the
final states as required for bosons.  The elastic differential
cross section is
\begin{equation}
{d\sigma \over dt} = {|{\cal M}|^{2} \over 16 \pi s(s-4m_{\pi}^{2})}.
\label{eq:dsdt}
\end{equation}
It has qualitatively the right
behavior:  for energies just below the $\rho$ resonance the angular
distribution is forwardly peaked, for $\sqrt{s}$ near $m_{\rho}$, the
distribution becomes forward-backward symmetric, and finally above the
resonance it is backward dominated.  We can make a detailed comparison with
data by integrating Eq.~(\ref{eq:dsdt}) in order to get $\sigma_{el} (s)$.
The scalar and tensor coupling constants and widths are adjusted so
as to give resonant peaks which match experiment.  Their
numerical values along with the usual parameters for the $\rho$ are
$g_{\sigma}m_{\sigma}=1.85$,
$m_{\sigma}=0.525$ GeV, $\Gamma_{\sigma}=0.100$ GeV,
$g_{\rho}=6.15$, $m_{\rho}=0.775$ GeV, $\Gamma_{\rho}=0.155$ GeV,
$g_{f}m_{f}=7.2$, $m_{f}=1.274$ GeV and
$\Gamma_{f}=0.176$ GeV.  The results
are shown in Fig.~\ref{pipidata} along with experimental data from
Refs.~\cite{vs76} and ~\cite{sdp73}.  We obtain superb agreement with
experiment for the energies shown.  Other $\pi\pi$ processes
($\pi^{\pm}\pi^{0} \rightarrow \pi^{\pm}\pi^{0}$) proceed analogously.  With
all three cross sections, we can compute the rate of pair production
using Eqs.~(\ref{eq:sigab}) and (\ref{eq:hatsigma}), which
is exact within the soft photon approximation.  In Fig~\ref{compirate}
we show this rate along with the former (approximate) result of
Eq.~(\ref{eq:sigpipi}).  The net result is a suppression from the approximate
result by a mere factor $\sim$ 2.  Through this comparison
we notice that $e^{+}e^{-}$ pairs with masses $\stackrel{<}{\sim}$ 100 MeV
are produced in $\pi\pi$ scattering at any kinematically allowed $\sqrt{s}$
and are therefore dominated near the $\rho$-peak;
whereas higher mass pairs ($M \approx$ 300 MeV) are effectively produced
through $\pi\pi$ scattering with $\sqrt{s} \stackrel{>}{\sim} m_{\rho}$.
When $M \approx 300$ MeV in Eqs.~(\ref{eq:sigab}) and (\ref{eq:genrate})
with $\sqrt{s}$ near the kinematically determined minimum (just below
the $\rho$-peak) the production rate is strongly suppressed by the
logarithm and radical.  Admittedly the soft photon approximation is less
reliable for high masses but this suggests that
it is imperative to have a reasonably accurate
calculation of the cross section for $\sqrt{s} > m_{\rho}$.   Such a
calculation would be incomplete if it excluded the tensor $f(1270)$ meson.

\section{Quark processes}
\label{sec:quarkbrems}

Quark scattering processes introduce additional difficulties due
to their fractional charge, their color, and their possibility for
elastic scattering with their own force mediating bosons, the gluons.
But we have developed the formalism to handle the general case.  The
same collection of diagrams from
Fig.~\ref{generice+e-} is considered here with the charged external lines
being $u$ or $d$ quarks (or antiquarks) and neutral external lines being
gluons.  In total we include six quark-quark (or antiquark) and four
quark-gluon (or antiquark) diagrams.  Other processes like $q\bar q
\rightarrow gg$ and its reverse process turn out to give a rather small
contribution to the $e^{+}e^{-}$ production rate
(of order 8 \% of our chosen processes) so we choose to neglect them here.
It should also be mentioned that the approximate $|\epsilon \cdot J|^{2}$
cannot be applied to these.
The electromagnetic amplitude for the partonic processes
included is obtained from Eq.~(\ref{eq:ejsquared}) by
setting the masses to $m_{q}$ and by setting $Q_{a}$ and $Q_{b}$ to the
appropriate quark electric charges.  The strong interaction differential
cross sections $d\sigma_{qq}/dt$ and $d\sigma_{qg}/dt$ are well known
in the perturbative vacuum at the one-gluon-exchange level to be~\cite{jl82}
\begin{equation}
{d \sigma_{a b} \over dt} = {C_{ab} 2\pi \alpha_{s}^{2} \over t^2},
\label{eq:dsigabvac}
\end{equation}
where
\begin{equation}
C_{ab} = \left\lbrace {1 {\ (qg\rightarrow \ qg)} \atop {4 \over 9}
{\ (qq\ \rightarrow \ qq)}} \right. .
\end{equation}
For hot hadronic matter this is clearly inadequate.  Let us consider
the dominant $t$-channel gluon exchange amplitude for the diagram
in Fig.~\ref{gluonxchng}.  ${\cal M}^{ab} = \alpha_{s} \Gamma_{\mu}^{a}
{\cal D}^{\mu \nu} \Gamma_{\nu}^{b}$, where ${\cal D}^{\mu \nu} =
g^{\mu \nu}/t$ is the gluon propagator and $\Gamma_{\mu,\nu}^{a,b}$
are vertex functions for particles {\em a,b}.  It is obvious that
${\cal M}^{ab}$ is singular in the zero momentum transfer limit.  But
at finite temperatures static color-electric fields are shielded by
quarks and gluons in the plasma, and color-magnetic fields are also
shielded by non-perturbative effects~\cite{pd85}.  Color-electric
shielding modifies the $\mu = \nu = 0$ components of ${\cal D}^{\mu \nu}$
in such a way that
\begin{equation}
{\cal M}_{E}^{ab} \simeq \alpha_{s}(t) \Gamma_{0}^{a}{\cal D}^{00}
\Gamma_{0}^{b} \simeq {\alpha_{E}(t) \over t} \Gamma_{0}^{a}
\Gamma_{0}^{b},
\end{equation}
where the running coupling constant~\cite{pd85}
\begin{equation}
{\alpha_{E}(t) \over t} = {\alpha_{s}(t) \over
 t-m_{E}^{2}},
\end{equation}
and the color-electric mass $m_{E}^{2} = 6\pi\alpha_{s}T^{2}$. For a
medium-modified magnetic scattering amplitude we obtain
\begin{equation}
{\cal M}_{M}^{ab} \simeq \alpha_{s}(t) \Gamma_{i}^{a}{\cal D}^{ij}
\Gamma_{j}^{b} \simeq -{\alpha_{M}(t) \over t} \Gamma_{i}^{a}
\Gamma_{i}^{b},
\end{equation}
where the effective magnetic coupling ${\alpha_{M}(t) / t}=
{\alpha_{s}(t) / (t-m_{M}^{2})}$ and the color-magnetic mass
$m_{M}^{2} = c \alpha_{s}^{2} T^{2}$.  Lattice gauge
calculations~\cite{td82} give $c \simeq$ 20--30.  In the small
$t$ limit the vertex functions reduce to $\Gamma_{\mu}^{a}
\simeq p_{\mu}^{a}$, where $p_{\mu}^{a}$ is the momentum of particle
$a$.  Averaging over color and spin degrees of freedom gives
the differential cross section~\cite{pd85}
\begin{equation}
{d\sigma_{ab} \over dt} = C_{ab} {8\pi \over s^{2}t^{2}}
\left\lbrack \alpha_{E}(t)p_{0}^{a}p_{0}^{b} - \alpha_{M}(t)
\vec p^{\ a} \cdot \vec p^{\ b} \right\rbrack^{2},
\label{eq:dsabdt}
\end{equation}
where $s=(p_{a}+p_{b})^{2}$.

Unfortunately, $d\sigma_{ab}/dt$ depends on the quark-quark (gluon)
orientation in the scattering through
$\vec p_{a} \cdot \vec p_{b}$.  For arbitrary
orientation of vectors $\vec p_{a}, \vec p_{b}$ we can not obtain
a closed expression for $d\sigma^{ab}/dt$, but if vector $\vec p_{a}$
is antiparallel to $\vec p_{b}$, $d\sigma_{ab}/dt$ reduces to
\begin{equation}
{d\sigma^{ab} \over dt} = C_{ab} {\pi \over 2}
{ \alpha_{s}^{2}(t) (2t-m_{E}^{2}-m_{M}^{2})^{2} \over
(t-m_{E}^{2})^{2}(t-m_{M}^{2})^{2} }.
\label{eq:dsabdt2}
\end{equation}
Note that this expression is exact for massless quark (antiquark)
scattering.  For $T\sim$ 200--300 MeV, $\alpha_{s} \simeq$ 0.2--0.3,
$m_{E} \simeq$ 2.2$T$ and $m_{M} \simeq$ 1.5$T$.  If we assume that
the color-electric mass is roughly equal to the color-magnetic mass,
$m_{E} \simeq m_{M}$, then we can restore the $m_{q}$ dependence in
the differential cross sections
\begin{eqnarray}
{d\sigma^{qg} \over dt} &=&  {2\pi \alpha_{s}^{2}(t)
\over s^{2} (t-m_{E}^{2})^{2}  } (s-m_{q}^{2})^{2} \\
{d\sigma^{qq} \over dt} &=&  {4\over 9}{2\pi \alpha_{s}^{2}(t)
\over s^{2} (t-m_{E}^{2})^{2}  } (s-2m_{q}^{2})^{2}.
\label{eq:dsabdt3}
\end{eqnarray}
Using  these expressions we did not observe sizeable changes in the
dilepton production rates or final spectra.  So instead we use
Eq.~(\ref{eq:dsabdt2}) henceforth.
Note that in the limit $m_{E},m_{M} \rightarrow 0$ (free space), we
reproduce Eq.~(\ref{eq:dsigabvac}) for the perturbative vacuum.

Putting together Eqs.~(\ref{eq:pairprod}), (\ref{eq:ejsquared}),
(\ref{eq:genrate}) and (\ref{eq:dsabdt2}), we have the rate
of $e^{+}e^{-}$ production from all the quark(antiquark)-gluon
or quark(antiquark)-quark processes.
In Figs.~\ref{approxqrates} and ~\ref{quarkrates}
we compare usage of $|\epsilon \cdot J|^{2}_{approx}$ to usage
of $|\epsilon \cdot J|^{2}_{exact}$ in calculating the rates
of $e^{+}e^{-}$ production from the sum of $qq (q\bar q)$ and
$qg (\bar qg)$ scattering with virtual bremsstrahlung at
temperatures $T=200$ and 300 MeV.  Three quark curves
are shown: current quarks, medium quarks,
and constituent quarks with masses $m_{q}=300$ MeV.
A remarkable feature emerges in the comparison between the correct and
approximate rates from current quarks: there is a huge suppression.
To perhaps better understand this effect
we present in Fig.~\ref{edjrat} the ratio
$R\equiv |\epsilon \cdot J|^{2}_{exact} / |\epsilon \cdot J|^{2}_{approx}$
for current, medium and constituent $u$ and $d$ quark scattering at
$\sqrt{s} = 775$ MeV.  Owing to the small mass of the current quarks,
the resulting ratio is $R=10^{-2}$--$10^{-3}$.  Another interesting
feature for current quarks is that  $R \rightarrow 1$ as
$\cos\theta \rightarrow 1$.  Even at these (relativistic) energies, the
electrodynamics in small-angle scattering is described very well by the
approximate amplitude.  However, these analyses invalidate
Eq.~(\ref{eq:approxejsquared}) for high momentum-transfer current quark
scattering at these energies.  Yet, for medium and constituent
quarks the ratio $R \sim 1$, so for scattering of these particles the
approximate expression, Eq.~(\ref{eq:approxejsquared}) is rather good.
Fig.~\ref{edjrat} also includes a curve for $\pi^{+}\pi^{-}$
scattering, where again the approximate electromagnetic factor
is quite near the exact result.

Since the quark mass dependence in the production rate is roughly
$1/m_{q}^{2}$ and the temperature dependence is roughly
$T^{6}$, an upper limit on the rate is obtained using
$m_{q}=5$ MeV at
$T=300$ MeV.  Similarly, a lower limit on the rate is obtained using
$m_{q}=300$ MeV at the phase transition temperature.  Based on the
results of Fig.~\ref{quarkrates} we conclude that at $T=200$ MeV
the (soft) virtual photon emissivities from competing
medium quark versus pionic processes in the mass region $M\stackrel{<}{\sim}
2m_{\pi}$ differ by less than a factor 4.  In a different energy regime
Kapusta et al.~\cite{jkplds91} reached the same conclusion
while comparing real (energetic) photon emissivities of the
QGP to that of hadronic matter.  This
equal-luminous property appears also in the annihilation channels
$q\bar q \rightarrow e^{+}e^{-}$ versus $\pi^{+}\pi^{-}\rightarrow
\rho \rightarrow e^{+}e^{-}$ but only after an (invariant mass) integrated
total emissivity.  In particular,
the emissivities $dN^{e^{+}e^{-}}/d^{4}x$ at $T=200$ MeV for pions, medium
quarks, and constituent quarks are approximately
$\simeq 2.5, 1.2$ and $\simeq 1.1 (\times 10^{-9} {\rm GeV}^{4})$,
respectively.

\section{Zero (total) momentum pairs}
\label{sec:zeromomentum}

As was pointed out by Kapusta~\cite{jik89}, the production rate for soft
dileptons could differ by orders of magnitude from the naive prediction
of quark-antiquark annihilation.  The enhancement might be due to a
modification in the quark dispersion relation in thermalized hadronic matter.
Furthermore, he argued that the zero (total) momentum $e^{+}e^{-}$ mass
spectrum is a direct measure of the quark dispersion relation and hence of
chiral symmetry.  A few possibilities for modifications are as follows.
First, ignoring interactions with the matter would lead to a
light-quark dispersion relation $\omega_{1} = k$.  Secondly,
interactions with the plasma would result in a
medium quark dispersion relation $\omega_{2} = (k^{2} + T^{2})^{1/2}$.
Still another possibility is to use weak-coupling methods
self-consistently~\cite{jk82}, where one possible parametrization is
$\omega_{3} = (1/3+k/T)k/(1+k/T)$.  Having a modified relation between energy
and momentum for the quarks leads to a modification in the annihilation
rates for producing zero momentum lepton pairs.  We can
compare these rates with our soft photon approximation only at a
qualitative level since one calculation includes
scattering with bremsstrahlung and the other includes quark-antiquark
annihilation.  Nevertheless, it is interesting to see how they relate.
Another result with which we might compare more quantitatively
is one which uses the resummation techniques of Braaten and
Pisarski~\cite{ebrp90}.  They were
recently applied~\cite{ebrpty90} to a calculation of the
zero-momentum soft dilepton production rate in a QGP.  Here
the differential rate for producing pairs is related to discontinuities in
the photon self-energy.
We refer the reader to Ref.~\cite{ebrpty90} for further details and
merely state the results.
The partial rate from the annihilation and decay of soft quarks and
antiquarks exhibits structure: sharp peaks due to Van Hove
singularities.  In addition to the pole contributions, there are terms
resulting from cuts in the effective quark propagator.  The square
of a coherent sum of such contributions appears in the final answer.
The physical processes which give the pole-cut and cut-cut contributions
involve soft and hard quark-gluon scattering processes, which we have
considered in our bremsstrahlung calculation.  At small invariant mass, the
pole-cut term grows like $\sim 1/M^{2}$ and the cut-cut term grows like
$\sim 1/M^{4}$; they completely overwhelm the structure due to
the pole-pole term.

In order to compare our result
with either calculation mentioned, we must restrict the lepton pair to
have zero total momentum $\vec q = 0$ in the rest frame of the matter.  With
such restrictions in Eq.~\ref{eq:pairprod}, we cast the cross section
into the form
\begin{equation}
{d^{4}\sigma^{e^{+}e^{-}} \over dMd^{3}q} (\vec q = 0)
= {\alpha^{2} \over 4\pi^{3}}  {\widehat \sigma(s) \over M^{4}},
\end{equation}
where the weighted elastic cross section $\widehat \sigma(s)$ is that
which we have defined in Eq.~\ref{eq:hatsigma}.  The rate
$dN/d^{4}xdMd^{3}q$ $(\vec q=0)$ is then obtained by
replacing $d\sigma/dM^{2}$ with $d^{4}\sigma/dMd^{3}q$ $(\vec q=0)$ in
Eq.~\ref{eq:epemprodrate}.  In Fig.~\ref{resumm} we present the comparison
of our (soft photon) approximate results, a parametrization
of the results of Braaten, Pisarski and Yuan obtained using resummation
techniques in QCD perturbation theory,  along with the rate
using the quark dispersion relation $\omega_{3} = (1/3+k/T)k/(1+k/T)$
in the $q\bar q$ annihilation contribution. The other dispersion relations
give even smaller rates. For dilepton masses $M \stackrel{<}{\sim}$ 0.1 GeV,
the production rate in the soft photon approximation is very near the result
from QCD perturbation theory and much larger than the
rate using the modified quark dispersion relation.  Of course,
the latter only includes annihilation whereas, the first two
include scattering--so there is a qualitative difference.  The low-mass
dependence in the rates is $\sim 1/M^{4}$ both in the perturbation theory
result as well as the soft photon approximation.  This dependence is
due to the processes of scattering hard quarks on hard gluons~\cite{ebrpty90},
which we have taken into account in the soft photon calculation.  For masses
0.1 $\stackrel{<}{\sim} M \stackrel{<}{\sim}$ 0.3 GeV, the $\sim 1/M^{2}$
dependence starts to become significant in the perturbation theory
result.  As previously mentioned, this dependence is due to the
pole-cut contributions; or physically, with scattering of soft quarks on
hard gluons.  These processes can only be included in our scheme
by retaining the next-to-leading term in the soft photon approximation.  The
matrix element $M^{\mu}$ in a soft photon approximation is expanded in powers
of the photon momentum $q$ as
\begin{equation}
M^{\mu} = {A^{\mu} \over q} + B^{\mu} + {\cal O}(q),
\end{equation}
where both $A^{\mu}$ and $B^{\mu}$ may calculated from knowledge of
the physical, i.e. on-shell $T$-matrix~\cite{lhproc}.  The term
$A^{\mu}/q$ gives the $1/M^{4}$ dependence in $dN/d^{4}xd^{3}q$ $(\vec q=0)$,
while $B^{\mu}$ is responsible for next-to-leading order terms.  In
particular only $B^{\mu} \ne 0$ gives rise to $1/M^{2}$ dependence.
The soft photon approximation with leading term only is quite close
to the (resummed) perturbation theory result
for masses $M\stackrel{<}{\sim}$ 0.1 GeV.  Most
likely, the next-to-leading term (added coherently) would enhance
the soft photon result for masses 0.1 $\stackrel{<}{\sim} M
\stackrel{<}{\sim}$ 0.3 GeV by a factor of 2--6.  However, such
good agreement between our calculation and that of Ref.~\cite{ebrpty90}
does two things.  It shows that scattering with virtual
bremsstrahlung (rather than annihilation or Compton-like processes) accounts
for most of the low-mass QGP-driven pairs and secondly, it lends support to
the validity of the soft-photon approximation as applied to quark
processes.

\section{Total electron-positron yield}
\label{sec:evolut}

In the previous sections we calculated the rates of dilepton production
by quark and pion scattering with virtual bremsstrahlung, i.e. the
number of electron-positron pairs produced per unit four-volume
$d^4x$ and with mass $M$ while being locally thermalized at temperature
$T$.  In order to obtain the spectrum with which to make an
experimental comparison, it is necessary to integrate over the time evolution
of the nuclear system.  Such an integration is possible only after
modeling the evolution in some way.  We shall make the standard
assumptions~\cite{jb83} about isentropic expansion and similarity flow in
the central region ($y=0$) of the heavy ion collision.  Under such
assumptions the
initial temperature and time can be related to the hadron distribution in
the final phase (for $T_{i}>T_{c}$) by
\begin{equation}
T_{i}^{3}\tau_{i} = {c\over 4a} {1\over \pi R_{A}^{2}}
{dN^{AA\rightarrow \pi} \over dy},
\label{eq:multiplic1}
\end{equation}
where $c \simeq 3.6$, $a\simeq 5.25$, $R_{A}$ is the nuclear radius
($R_{A} = 7.1$ fm for Pb), and $\tau_{i} \simeq$ 1 fm/c is the initial
thermalization time.  For LHC and RHIC energies the expected multiplicities
$dN^{AA\rightarrow \pi}/dy \simeq$ 3000 correspond to initial
temperature $T_{i} \simeq 300$ MeV.  As for the temperature of
the phase transition $T_{c}$, the lattice estimations provide a
range of $T_{c} \simeq$ 150--200 MeV.  We assume that the phase
transition is first-order, so at $T_{c}$ the mixed phase may be realized.
For possible decay temperatures we use $T_{f} = 140$ and $100$ MeV.  The
production in the pure quark and hadronic phases can be
calculated~\cite{kk86} from
\begin{equation}
{dN \over dydM} = 3\pi R_{A}^{2} T_{i}^{6}\tau_{i}^{2}
\int\limits_{T_{c}}^{T_{i}} {dT \over T^{7}} {dN^{q} \over d^{4}xdM},
\nonumber
\end{equation}
and
\begin{equation}
{dN \over dydM} = 3\pi R_{A}^{2} T_{i}^{6}\tau_{i}^{2}r^{2}
\int\limits_{T_{f}}^{T_{c}} {dT \over T^{7}} {dN^{\pi} \over d^{4}xdM},
\label{eq:multiplic2}
\end{equation}
while the mixed phase contribution from quark and pion processes
are
\begin{equation}
{dN^{q}_{mixed} \over dydM} = {\pi R_{A}^{2}\over 2}
\left({T_{i}\over T_{c}}\right)^{6} \tau_{i}^{2}(r-1)
{dN^{q} \over d^{4}xdM}(T=T_{c}),
\end{equation}
and
\begin{equation}
{dN^{\pi}_{mixed} \over dydM} = {\pi R_{A}^{2}\over 2}
\left({T_{i}\over T_{c}}\right)^{6} \tau_{i}^{2}r(r-1)
{dN^{\pi} \over d^{4}xdM}(T=T_{c}).
\end{equation}
Here $r$ denotes the ratio of the number of degrees of freedom
of the QGP constituents to that of the hadron gas constituents
($r\simeq 12$).

In Fig.~\ref{totalyield} we present the resulting invariant mass spectra
through virtual bremsstrahlung from quarks and pions using relevant
temperatures $T_{i}=300$, $T_{c}=200$ and $T_{f}=140$ MeV.  For this
collection of temperatures the pion cooling phase contribution is the larger
contributor to its final spectrum, whereas the mixed phase dominates
the resulting quark driven spectrum.  Note the difference between the
results here compared to our initial estimates in Fig. 4 of
Ref.~\cite{kh92}.  The suppression
is due in part to the four-momentum conservation at the photon-dilepton
vertices in Eq.~(\ref{eq:pairprod}) and to the phase-space correction
shown in Eq.~(\ref{eq:phasespace}). Also included in Fig.~\ref{totalyield}
is an estimate of the
background due to $\pi^{0}$ and $\eta$ Dalitz decay $\pi^{0}
\rightarrow e^{+}e^{-} + \gamma$, $\eta \rightarrow e^{+}e^{-}\gamma$.
We calculate these backgrounds using formulae from Ref.~\cite{jc91}.
This comparison suggests that $\pi^{0}$ Dalitz is the strongest source
of pairs having invariant mass less than the pion mass
and that $\eta$ Dalitz is the strongest source of
$e^{+}e^{-}$ pairs with invariant mass larger that $m_{\pi}$ but less
2$m_{\pi}$. For masses larger than 2$m_{\pi}$,
$\pi \pi$ annihilation will of course becomes dominant.
Superimposed on Fig.~\ref{totalyield} is also the yield
from  $q\bar q \rightarrow \gamma^{*} \rightarrow
e^{+}e^{-}$ annihilation in the Born approximation~\cite{kk86,jc91}.
Throughout most of the range of dilepton invariant mass that we consider,
the contribution from pion scattering with virtual bremsstrahlung and
from $\pi^{0}$, $\eta$ Dalitz decays are two or three orders of magnitude
above the $q\bar q$ annihilation spectrum.  Perturbative corrections
to this annihilation spectrum for finite $p_{\bot}$
are extremely important~\cite{swong92}.
An advantage of our soft photon approach is that
we could easily evaluate the bremsstrahlung contribution at finite
$q_{\bot}$ or even impose some cuts in transverse momentum.

Another quite reasonable possibility for the relevant temperatures
of the colliding
system is $T_{i}=250$, $T_{c}=150$ and $T_{f}=100$ MeV, corresponding
to those used in Ref.~\cite{jc91} in a recent study of low-mass
dilepton production.  We reproduce their $\pi^{\pm}\pi^{0}$ results
using our $|\epsilon \cdot J|^{2}_{approx}$ and
using the same parametrization of the total $\pi \pi$
cross section as did they.  In Fig~\ref{totalyield2} we show our
{\em complete} calculation using these temperatures.
Our final results are rather different from Ref.~\cite{jc91}
since we include several improvements: we have other pion processes, we
strictly conserve four-momentum at the photon-dilepton vertices, we
use the exact electromagnetic factor
and finally, we correct phase space to include the momentum
of the virtual photon.  It should also be mentioned that
the pion results of Ref.~\cite{jc91} should not
be taken seriously for invariant masses larger than $\sim 2m_{\pi}$.  They
do not include any $\pi\pi$ resonances in the cross section with masses
greater than the $\rho$ and
most importantly they do not correct phase space which is
essential for higher invariant masses.
For comparison purposes, we show in Fig.~\ref{totalyield3}
still a different collection of temperatures having the
lowest $T_{c}$ which is compatible with lattice gauge theory
data--$T_{i}=200$, $T_{c}=150$, and $T_{f}=140$ MeV.
As one can see from Figs.~\ref{totalyield2}--\ref{totalyield3},
the medium quark yield is always lower than the yields from
$\pi^{0}, \eta$ Dalitz as well as pion scattering with bremsstrahlung.
This being true,
detection of quark degrees of freedom within this invariant mass
region will be difficult in future heavy-ion experiments.
Dalitz decays can (in principle) be distinquished from bremsstrahlung
and removed from analyses.  Distinguishing pairs produced through
pion processes from those produced through {\em medium} quark processes
will be difficult since total yields from both are nearly the same
in all temperature scenarios we investigated.
However, these experiments may
be very important with regard to the study of pion dynamics in a
thermalized system.  For initial temperatures $T_{i} > 350$ MeV
($dN/dy \stackrel{>}{\sim}$ 4000 for LHC), pion bremsstrahlung
is stronger than $\eta$ Dalitz in the mass region 100
$\stackrel{<}{\sim} M \stackrel{<}{\sim}$ 300 MeV, but it is
weaker than $\pi^{0}$ Dalitz for masses less than 100 MeV.
 From Eqs.~(\ref{eq:multiplic1}--\ref{eq:multiplic2}), we see
the dilepton yield from thermalized pions is proportional
to the square of the hadron multiplicity $(dN/dy)^{2}$ whereas,
it is directly proportional to the multiplicity for
Dalitz decays.  Therefore, we expect that when the density of charged
particles $dN/dy \stackrel{>}{\sim}$ 4000, the yield of
$e^{+}e^{-}$ pairs with 100 $< M <$ 300 MeV will be
$\sim (dN/dy)^{2}$, but with $M \stackrel{<}{\sim}$ 100 MeV
it will be $\sim (dN/dy)$.

\section{Summary}
\label{sec:concl}

In a relativistic heavy-ion collision there is likely born a
hot thermalized system.  We argued that one of
the best probes of such a system must surely be electromagnetic in
its nature.  In this study
we have used low-mass electron-positron
pairs as such a vehicle by investigating their origin, their
abundance and behavior during these collisions.
A soft photon approximation was used to compare hadronic to
partonic scattering with virtual bremsstrahlung.  In particular,
radiation of virtual
quanta during $\pi \pi$ scattering with all isospin possibilities
were discussed and compared with quark-quark(antiquark) and
quark(antiquark)-gluon scattering accompanied by virtual radiation.
We conclude that pions and medium quarks are roughly equally luminous at
$T_{c}$, severely crippling our ability to separate one from the other.

We compared our quark-driven production rate for zero momentum
low-mass dileptons to that obtained using QCD resummation techniques.
Our soft-photon-approximate result is very near the resummation
rate for masses less than 100 MeV and a factor of 2--6 less than
the resummation rate for masses
larger than 100 (but less than 300 MeV).  We expect that the next-to-leading
term in the soft photon approximation (added coherently) would enhance
our result for masses 0.1 $\stackrel{<}{\sim} M \stackrel{<}{\sim}$ 0.3
GeV by 2--6.  Such good agreement for masses less than 100 MeV between
our soft virtual photon calculation and that of Ref.~\cite{ebrpty90}
using resummation techniques lends support to the validity of the soft photon
approach to this problem.
Using a (1+1) dimensional Bjorken hydrodynamic model to describe the
expansion of the system, we calculated $e^{+}e^{-}$ total yields for
different collections of temperatures $T_{i}, T_{c}$ and $T_{f}$.
For all reasonable sets of temperatures the constituent and the
medium quark-driven total dilepton yields are lower than $\pi^{0}$ and
$\eta$ Dalitz and lower than the pion-driven yield.  We must conclude that
it will be difficult to observe quark degrees of freedom (DOF) in this
region of soft dileptons at LHC or RHIC.  At the same time, it is important
to keep in mind that our quark-driven soft-photon approximate results
were below the pertubation theory results (shown in Fig.~\ref{resumm})
for invariant masses near $2m_{\pi}$ by a factor $\sim 4$.  This
precious factor of 4 may be enough to allow reasonably clear
observation of these DOF.  Another useful observation
is that for $T_{i}>$ 350 MeV ($dN/dy \stackrel{>}{\sim} 4000$
for LHC) the pion-driven production rate is larger than $\eta$ Dalitz
for masses 0.1 $\stackrel{<}{\sim} M \stackrel{<}{\sim}$ 0.3 GeV, though
still smaller than $\pi^{0}$ Dalitz for masses smaller than 0.1 GeV.
Therefore, we predict that for $dN/dy \stackrel{>}{\sim} 4000$ the
yield of dileptons at 0.1 $\stackrel{<}{\sim} M \stackrel{<}{\sim}$ 0.3
GeV will be proportional to $(dN/dy)^{2}$ which reflects the collective
effects in the thermalized pion gas.  Whereas, the dilepton
yield for $M \stackrel{<}{\sim}$ 0.1 GeV is proportional to
$(dN/dy)$, i.e. $\pi^{0}$ Dalitz dominated.  Future
heavy ion experiments (RHIC, LHC) will be useful for investigations
of pion dynamics at high densities with regard to their soft dilepton
mass spectra.

In our discussion of pion processes we assumed that the
$\pi\pi$ differential cross section did not change from its
free-function at finite temperatures.
In particular, we used the same expression for
$d\sigma/dt$ and therefore $\sigma_{el}$
near $T=T_{c}$ as we did for $T=0$.  It is clear that many-body
effects introduce temperature dependences into the meson masses, widths
and coupling constants.  Therefore,  the $\pi\pi$ elastic cross section will
likely be quite different near $T_{c}$.  In order to estimate the temperature
dependence of $d\sigma_{el}/dt$ and of $\sigma_{el}$, we
use as a first approximation the chiral perturbation theory modifications of
$F_{\pi}$ and $\Gamma_{\rho}$ which are~\cite{jghl87}
\begin{equation}
F_{\pi}(T) = F_{\pi}(0)\left( 1 - {T^{2}\over 8F_{\pi}^{2}(0)}\right),
\label{eq:fpi}
\end{equation}
and
\begin{equation}
\Gamma_{\rho}(T) = {\Gamma_{\rho}(0) \over 1-T^{2}/4F_{\pi}^{2}(0)}.
\label{eq:gammarho}
\end{equation}
The zero temperature values are $F_{\pi}(0)\simeq 0.098$ GeV and
$\Gamma_{\rho}(0) =0.155$ GeV.
Since the pion decay constant changes, so does the coupling of
the $\rho$ to the pions because $g_{\rho \pi\pi} = m_{\rho}/\sqrt{2}F_{\pi}$.
The $\rho$-meson mass does not change appreciably with
temperature~\cite{cgjk91}.  We see from
Eqs.~(\ref{eq:fpi}) and ~(\ref{eq:gammarho}) that the contribution of
the non-resonance region ($\sqrt{s} \stackrel{<}{\sim} 0.6$ GeV) in
$\sigma_{el}$ will be enhanced with increasing $T$, whereas
the contribution from the $\rho$-resonace will be suppressed.  Basically,
the $\rho$-peak gets widened and reduced with increasing $T$.  It
gets completely melted at $T_{c}$.
The spectrum $dN^{e^{+}e^{-}}/dMdy$ obtained from Eq~(\ref{eq:genrate})
by some portion of a space-time integration for
bremsstrahlung during pion scattering is proportional to the
area under the $\sigma_{el}(s)$ curve.  Incorporating the
temperature dependences in this fashion we observe little or no change
in this area.  This translates into the fact that we do not observe
a large difference (less than a factor 2 enhancement)
in the production rates or total yield with a temperature dependent cross
section.  Finally, we did not consider
the change of $\Gamma_{\pi^{0}\rightarrow e^{+}e^{-}\gamma}$ or
$\Gamma_{\eta \rightarrow e^{+}e^{-}\gamma}$ because the lifetime of
the $\pi^{0}$ and the $\eta$ is such that only very few will
decay during the mixed or hadronic phases.  So the pion bremsstrahlung
contribution to the dilepton spectra is not very sensitive to
temperature effects, if temperature dependences are well
accounted by Eqs.~(\ref{eq:fpi}) and ~(\ref{eq:gammarho}).

One topic somewhat ignored in our study is the transverse momentum
dependence of the produced $e^{+}e^{-}$ pairs.  Apart from
the zero-momentum comparison we made, we presented all the results
integrated over transverse momentum.  Mostly because this study
represents a first detailed analysis of quark and
pion bremsstrahlung.  There
are at least two reasons we might benefit from transverse momentum
analyses.  Firstly,
presentation of experimental data would mostly likely be
$q_{\bot}$-restricted due to detector (in)efficiencies and secondly
there seems to be the possibility that Dalitz contributions
might be suppressed more than would be bremsstrahlung for some
restricted values of $q_{\bot}$~\cite{khvecg92}.  Finally, note
that the soft photon approach that we have taken to this low-mass pair
production problem can be equally well applied to the production of soft
real photon production.

\acknowledgements

One of us (V.E.) is grateful to Profs. S. K. Mark and P. Depommier for
organizing his stay at McGill University and grateful to the physics
department of McGill University for its warm hospitality.
We acknowledge useful discussions with J. I. Kapusta.  This work
was supported in part by the Natural Sciences and Engineering Research
Council of Canada and by the FCAR fund of the Qu\'ebec Government.

\appendix{Electromagnetic factor $|\epsilon \cdot J|^{2}$}
\label{app:edotjcorr}

In this appendix we present the details for evaluating the exact
electromagnetic factor for the soft-photon approximation.
Recall the current in the scattering process $ab \rightarrow cd$:
\begin{equation}
J^{\mu} = -Q_{a} {p_{a}^{\mu} \over p_{a}\cdot q} -
Q_{b} {p_{b}^{\mu} \over p_{b}\cdot q} +
Q_{c} {p_{c}^{\mu} \over p_{c}\cdot q}
+ Q_{d} {p_{d}^{\mu} \over p_{d}\cdot q}.
\label{eq:jmuappx}
\end{equation}
Summing over polarizations aided by current
conservation leads to \cite{rf49}
\begin{equation}
\sum\limits_{\rm pol} \epsilon_{\mu} \epsilon_{\nu}
J^{\mu} J^{\nu} = - J^{2}.
\end{equation}
The particles' initial velocities are $\vec \beta_{a}, \vec \beta_{b}$
and final velocities are $\vec \beta_{c}, \vec \beta_{d}$
($c=1$).  We move to the center-of-velocity frame of any
charged participants where in the soft-photon limit
$|\vec \beta_{i}| = \beta$ for all $i$, $\vec \beta_{a} \cdot \vec\beta_{b}
= \vec \beta_{c} \cdot \vec\beta_{d} = -\beta^{2}$
and $\vec \beta_{a} \cdot \vec
\beta_{c} = \vec \beta_{b} \cdot \vec
\beta_{d} = \beta^{2}\cos \theta$.
Since $p_{i}^{2} = m^{2}_{i}$, we have
\begin{eqnarray}
|\epsilon \cdot J|^{2} &=& -{Q_{a}^{2}m_{a}^{2} \over E_{a}^{2}q_{0}^{2}
(1-\vec \beta_{a} \cdot \vec n)^{2}}
-{Q_{b}^{2}m_{b}^{2} \over E_{b}^{2}q_{0}^{2}
(1-\vec \beta_{b} \cdot \vec n)^{2}}
-{Q_{c}^{2}m_{c}^{2} \over E_{c}^{2}q_{0}^{2}
(1-\vec \beta_{c} \cdot \vec n)^{2}} \nonumber\\
& \ & -{Q_{d}^{2}m_{d}^{2} \over E_{d}^{2}q_{0}^{2}
(1-\vec \beta_{d} \cdot \vec n)^{2}}
-{2Q_{a}Q_{b}(1-\vec \beta_{a} \cdot \vec \beta_{b})
\over q_{0}^{2}(1-\vec \beta_{a} \cdot \vec n)
(1-\vec \beta_{b} \cdot \vec n)}
+{2Q_{a}Q_{c}(1-\vec \beta_{a} \cdot \vec \beta_{c})
\over q_{0}^{2}(1-\vec \beta_{a} \cdot \vec n)
(1-\vec \beta_{c} \cdot \vec n)} \nonumber\\
& \ &+{2Q_{a}Q_{d}(1-\vec \beta_{a} \cdot \vec \beta_{d})
\over q_{0}^{2}(1-\vec \beta_{a} \cdot \vec n)
(1-\vec \beta_{d} \cdot \vec n)}
+{2Q_{b}Q_{c}(1-\vec \beta_{b} \cdot \vec \beta_{c})
\over q_{0}^{2}(1-\vec \beta_{b} \cdot \vec n)
(1-\vec \beta_{c} \cdot \vec n)} \nonumber\\
& \ &+{2Q_{b}Q_{d}(1-\vec \beta_{b} \cdot \vec \beta_{d})
\over q_{0}^{2}(1-\vec \beta_{b} \cdot \vec n)
(1-\vec \beta_{d} \cdot \vec n)}
-{2Q_{c}Q_{d}(1-\vec \beta_{c} \cdot \vec \beta_{d})
\over q_{0}^{2}(1-\vec \beta_{c} \cdot \vec n)
(1-\vec \beta_{d} \cdot \vec n)}
\label{eq:edotjfinal}
\end{eqnarray}
where $\vec n \equiv \vec q/|\vec q \/|$ is the unit vector for the photon.
We are interested in an angular average over the emitted photon's direction
\begin{equation}
\int {d\Omega_{q} \over 4 \pi}
|\epsilon \cdot J|^{2}.
\end{equation}
Carrying out the integrals for the first four terms in
Eq.~(\ref{eq:edotjfinal})
is elementary.  The result for each is just the charge squared over
$q_{0}^{2}$.  The remaining terms are readily evaluated using a
two-parameter Feynman integral
\begin{equation}
{1\over ab} = \int\limits_{0}^{1} {dx \over [ax + b(1-x)]^{2}}.
\end{equation}
We show, for example, integration of the fifth term in
Eq.~(\ref{eq:edotjfinal}).
\begin{eqnarray}
& \ & -2Q_{a}Q_{b}(1+\beta^{2})\int {d\Omega_{q} \over 4\pi}
{1 \over q_{0}^{2}
(1-\vec \beta_{a} \cdot \vec n)(1-\vec \beta_{b} \cdot \vec n)}
\nonumber\\
&=&-2Q_{a}Q_{b}(1+\beta^{2})\int\limits_{0}^{1} dx \int {d\Omega_{q}
\over 4\pi}
{1 \over q_{0}^{2}
\left[1- \vec n \cdot \left(\vec \beta_{a}x+\vec \beta_{b}(1-x)
\right)\right]^{2}}
\nonumber\\
&=&-2Q_{a}Q_{b}(1+\beta^{2})\int\limits_{0}^{1} dx
{1 \over q_{0}^{2}
\left(1- \left|\vec \beta_{a}x+\vec \beta_{b}(1-x)\right|^{2}\right)}
\nonumber\\
&=&-2Q_{a}Q_{b}(1+\beta^{2})\int\limits_{0}^{1} dx
{1 \over q_{0}^{2}(- 4\beta^{2} x^{2}+ 4\beta^{2} x + 1-\beta^{2})}
\nonumber\\
&=&{-2Q_{a}Q_{b}(1+\beta^{2}) \over -4\beta^{2}} \int\limits_{0}^{1} dx
{1 \over q_{0}^{2}\left( x-{1\over 2}+
\sqrt{{1\over 4} + {1-\beta^{2} \over 4\beta^{2}}} \ \right)
\left( x-{1\over 2}-
\sqrt{{1\over 4}  + {1-\beta^{2} \over 4\beta^{2}}} \ \right)} \nonumber\\
&=&{-2Q_{a}Q_{b}(1+\beta^{2}) \over 4\beta q_{0}^{2}} \int\limits_{0}^{1} dx
\left\lbrace {1 \over  x-{1\over 2}+
\sqrt{{1\over 4} + {1-\beta^{2} \over 4\beta^{2}}}}
+ {1 \over x-{1\over 2}-
\sqrt{{1\over 4}  + {1-\beta^{2} \over 4\beta^{2}}}} \right\rbrace
\end{eqnarray}
The
integration is then trivial; yielding,
\begin{equation}
=-2Q_{a}Q_{b}(1+\beta^{2}) {1 \over 2q_{0}^{2}\beta}
\ln \left( {1+\beta
\over 1-\beta}\right).
\end{equation}
Evaluation of all other terms in Eq.~(\ref{eq:edotjfinal}) follows
the same prescription.  The net result of such an exercise is
\begin{eqnarray}
|\epsilon \cdot J|^{2}
&=& \left. {1\over q_{0}^{2}} \right\lbrace
-( Q_{a}^{2} + Q_{b}^{2} + Q_{c}^{2} + Q_{d}^{2})
- 2(Q_{a}Q_{b}+Q_{c}Q_{d})
{1+\beta^{2} \over 2\beta} \ln \left(
{1+\beta \over 1-\beta} \right)
\nonumber\\
& \ & + 2(Q_{a}Q_{c} + Q_{b}Q_{d}) {\gamma (1-\beta^{2}\cos\theta)
\over \beta \sqrt{(1-\cos\theta)[\beta^{2}\gamma^{2}(1-\cos\theta)
+2]}}\nonumber\\
&\ & \ \ \ \ \times
\ln \left( {\sqrt{\beta^{2}\gamma^{2}(1-\cos\theta)+2} +
\sqrt{\beta^{2}\gamma^{2}(1-\cos\theta)} \over
\sqrt{\beta^{2}\gamma^{2}(1-\cos\theta)+2} -
\sqrt{\beta^{2}\gamma^{2}(1-\cos\theta)}}
\right) \nonumber\\
& \ & + 2(Q_{a}Q_{d} + Q_{b}Q_{c}) {\gamma(1+\beta^{2}\cos\theta)
\over \beta \sqrt{(1+\cos\theta)[\beta^{2}\gamma^{2}(1+\cos\theta)+
2]}} \nonumber\\
&\ & \ \ \ \ \times \left.
\ln \left( {\sqrt{\beta^{2}\gamma^{2}(1+\cos\theta)+2} +
\sqrt{\beta^{2}\gamma^{2}(1+\cos\theta)} \over
\sqrt{\beta^{2}\gamma^{2}(1+\cos\theta)+2} -
\sqrt{\beta^{2}\gamma^{2}(1+\cos\theta)}}
\right)  \right\rbrace
\label{eq:corredj3}
\end{eqnarray}
The velocity and angle can be replaced with invariants through
\begin{equation}
\beta = \sqrt{s-(m_{a}+m_{b})^{2} \over s-(m_{a}-m_{b})^{2}},
\label{eq:invs}
\end{equation}
and
\begin{equation}
\cos\theta = 1 + {2t \over s-(m_{a}+m_{b})^{2}}.
\label{eq:invt}
\end{equation}
In future calculations this general expression
will be very useful, but for our purposes here it is enough to use
equally massed particles, $m$.
Then Eq.~(\ref{eq:invs}) reduces to
\begin{equation}
\beta = \sqrt{{s-4m^{2} \over s}},
\end{equation}
and Eq.~(\ref{eq:invt}) reduces to
\begin{equation}
\cos\theta = 1 + {2t \over s-4m^{2}}.
\end{equation}
Carrying out these substitutions in order to express the result in
terms of invariants $s$ and $t$ we arrive at
\begin{eqnarray}
|\epsilon \cdot J|^{2}
&=& \left. {1\over q_{0}^{2}} \right\lbrace
-( Q_{a}^{2} + Q_{b}^{2} + Q_{c}^{2} + Q_{d}^{2}) \nonumber\\
&\ &- 2(Q_{a}Q_{b}+Q_{c}Q_{d})
{s-2m^{2} \over \sqrt{s(s-4m^{2})}} \ln \left(
{\sqrt{s} + \sqrt{s-4m^{2}} \over \sqrt{s} - \sqrt{s-4m^{2}}} \right)
\nonumber\\
& \ & + 2(Q_{a}Q_{d} + Q_{b}Q_{c}) {s-2m^{2}+t
\over \sqrt{(s+t-4m^{2})(s+t)}}
\ln \left( {\sqrt{s+t} + \sqrt{s+t-4m^{2}} \over \sqrt{s+t} -
\sqrt{s+t-4m^{2}}}
\right) \nonumber\\
& \ & \left. + 2(Q_{a}Q_{c} + Q_{b}Q_{d}) {2m^2-t \over \sqrt{t(t-4m^{2})}}
\ln \left( {\sqrt{4m^{2}-t} + \sqrt{-t} \over \sqrt{4m^{2}-t} - \sqrt{-t}}
\right)  \right\rbrace,
\label{eq:correctedotj}
\end{eqnarray}
where $\sqrt{s} \ge 2m$ and $-(s-4m^{2}) \le t \le 0$.

\appendix{Approximate electromagnetic factor}
\label{app:edotjapprox}

It is perhaps more convenient to have an approximate, yet reasonably
reliable expression for $|\epsilon \cdot J|^{2}$.  This appendix
is devoted to such a calculation.   Let $p^{\mu}_{a} =
(E_{a}, \vec p_{a})$ be the four-vector for particle $a$,
and similarly for particles $b,c$ and $d$. Let $q^{\mu} = q_{0}(1,\vec n)$
again
be the photon's space-time description and choose the radiation gauge so that
$\epsilon^{\mu} = (0, \vec \epsilon\ )$.
Then the quantity $|\epsilon \cdot J|^{2}$ can be written
\begin{equation}
|\epsilon \cdot J|^{2} ={1\over q_{0}^{2}} \left| \vec \epsilon \cdot
\left( {Q_{c} \vec \beta_{c} \over (1-\vec n \cdot \vec \beta_{c})}
- {Q_{a} \vec \beta_{a} \over (1-\vec n \cdot \vec\beta_{a})}
+ {Q_{d} \vec \beta_{d} \over (1-\vec n \cdot \vec\beta_{d})}
- {Q_{b} \vec \beta_{b} \over (1-\vec n \cdot \vec\beta_{b})} \right)
\right|^{2}.
\label{eq:approxedji}
\end{equation}
Let us work in the center-of-velocity frame of any charged
participants and write $\vec \beta_{a}
= \vec \beta$, $\vec \beta_{c} = \vec \beta + \Delta {\vec\beta}$,
$\vec\beta_{b}= -\vec\beta$ and
$\vec \beta_{d} = -(\vec \beta + \Delta \vec\beta)$.  We also assume
that $Q_{a}=Q_{c}$, $Q_{b}=Q_{d}$.  Then
if the scattering
is such that $\Delta \vec \beta$ is small relative to $\vec \beta$,
a good approximation to Eq. (\ref{eq:approxedji}) is
\begin{equation}
\simeq {1\over q_{0}^{2}} \left| \vec \epsilon \cdot
\left( {Q_{a}\left( \Delta \vec \beta + \vec n \times (\vec \beta \times
\Delta \vec \beta)\right)
 \over (1-\vec n \cdot \vec \beta)^{2}}
+ {Q_{b}\left(
-\Delta \vec \beta + \vec n \times (\vec \beta \times
\Delta \vec \beta \right)
 \over (1+\vec n \cdot \vec \beta)^{2}}\right)
\right|^{2}
\label{eq:approxedjii}
\end{equation}
where $\vec \beta$ is the initial (or average) velocity.  Since we
consider small angle deflection, $\Delta \vec \beta$ is
approximately perpendicular to the incident direction.  Next we
explicitly sum over polarizations by defining the unit vectors
$\hat \epsilon_{\|}$ and $\hat \epsilon_{\bot}$
to be the polarization vectors parallel and perpendicular to the
plane containing $\vec \beta$ and $\vec n$.  Averaging over
the azimuth of the scattered particles, the two polarization
summands become
\begin{equation}
|\epsilon \cdot J|^{2}_{\|} = {| \Delta \vec\beta |^{2} \over
2 q_{0}^{2}}
\int {d\Omega \over 4\pi}
 \left( {Q_{a} \over (1-\beta\cos\theta)} + {Q_{b} \over
(1+\beta\cos\theta)}   \right)^{2}
\end{equation}
and
\begin{equation}
|\epsilon \cdot J|^{2}_{\bot} = {|\Delta \vec\beta |^{2} \over
2q_{0}^{2}}
\int {d\Omega \over 4\pi}
\left( {Q_{a}(\beta-\cos\theta) \over (1-\beta\cos\theta)^{2}} +
{Q_{b} (\beta+\cos\theta) \over
(1+\beta\cos\theta)^{2}}   \right)^{2}.
\end{equation}
Integration of these expressions is elementary but quite
lengthy.  We perform the necessary work and add the contributions
from both polarizations to get
\begin{eqnarray}
\left| \epsilon \cdot J\right|^{2}_{approx}
&=& {|\Delta\vec \beta|^{2} \gamma^{2} \over q_{0}^{2}}
\left\lbrace {Q_{a}^{2}\over 6} + {Q_{b}^{2}\over 6} +
{(Q_{a}^{2}+Q_{b}^{2})\over 2 } +\beta{Q_{a}Q_{b}\over 2} -
{1\over \beta^{2}}{Q_{a}Q_{b}\over 2} \right. \nonumber\\
&\ & \left. + {Q_{a}Q_{b}\over 2\gamma^{2}} \left[
{1\over \beta} + {1\over 2\beta^{3}} + {\beta \over 2} \right]
\ln\left({1+\beta \over 1-\beta}\right) \right\rbrace.
\end{eqnarray}
For arbitrary masses $m_{a}=m_{c}$ and $m_{b}=m_{d}$,
$\gamma^{2}|\Delta \vec \beta|^{2}
= -t/(m_{a}m_{b})$ and
$\beta^{2} = \lbrace s-(m_{a}+m_{b})^{2}\rbrace /
\lbrace s-(m_{a}-m_{b})^{2} \rbrace$.
It is enough for our purposes here to use equally massed particles
$m_{a} = m_{b} = m$, in which case $\gamma^{2}|\Delta \vec \beta|^{2}
= -t/m^{2}$ and $\beta^{2} = (s-4m^{2})/s$.  Restoring the invariants
$s$ and $t$,
we get the final (approximate) result of
\begin{equation}
\left| \epsilon \cdot J\right|^{2}_{approx}
= {2 \over 3} {1\over q_{0}^{2}}
\left( {-t\over m^{2}}\right) \left[ ( Q_{a}^{2}
+ Q_{b}^{2}) -{3\over 2} Q_{a} Q_{b}f(s) \right],
\end{equation}
where
\begin{eqnarray}
f(s) & = & {s \over 2(s-4m^{2})} - {s-4m^{2}\over 2s}
\nonumber\\
& \ & -{m^{2} \over s} \left\lbrace
{2 \sqrt{s} \over \sqrt{s-4m^{2}}}
 + \left({\sqrt{s} \over \sqrt{s-4m^{2}}}\right)^{3} \right.
\nonumber\\
& \ &\left. + {\sqrt{s-4m^{2}}\over \sqrt{s}} \right\rbrace
\ln {\left( {\sqrt{s} + \sqrt{s-4m^{2}} \over
\sqrt{s} - \sqrt{s-4m^{2}}} \right)}.
\end{eqnarray}
Strictly speaking, this expression is valid only for $|t| < 4m^{2}$.

\appendix{Rate calculation from general scattering process}
\label{app:dne+e-d4xdm2}

The starting point for the rate calculation (from kinetic theory)
is
\begin{eqnarray}
{dN^{e^{+}e^{-}}_{ab}\over d^{4}x dM^{2}} &=& g_{ab}
\int ds\int {d^{3}p_{a}\over (2\pi)^{3}}\int {d^{3}p_{b}\over (2\pi)^{3}}
e^{-\beta(E_{a}+E_{b})} {d\sigma(s)\over dM^{2}}
v_{rel} \delta\left(s-(p_{a}+p_{b})^{2}
\right), \nonumber\\
\end{eqnarray}
where $g_{ab}=(2s_{a}+1)(2s_{b}+1)N_{c}^{a}N_{c}^{b}$ is the (spin and
color) degeneracy.
Trivial angular integrations can be done
\begin{eqnarray}
{dN^{e^{+}e^{-}}_{ab}\over d^{4}x dM^{2}} &=& {g_{ab}\over 32\pi^{4}}
\int ds\int dp_{a} \int dp_{b} {p_{a} p_{b}
e^{-\beta(E_{a}+E_{b})} \over E_{a}E_{b}}
{d\sigma_{ab}^{e^{+}e^{-}} \over dM^{2}}
\lambda^{1/2}(s,m_{a}^{2},m_{b}^{2}) \Theta(\chi)
\end{eqnarray}
where
\begin{equation}
\lambda(s,m_{a}^{2},m_{b}^{2}) =
{s\left(s-2[m_{a}^{2}+m_{b}^{2}]\right) + (m_{a}^{2}-m_{b}^{2})^{2}}
\nonumber
\end{equation}
and
\begin{equation}
\chi = \left[4E_{a}E_{b}(s-m_{a}^{2}-m_{b}^{2})-
(s-m_{a}^{2}-m_{b}^{2})^{2}-4m_{a}^{2}E_{b}^{2} -4m_{b}^{2}E_{a}^{2}
+4m_{a}^{2}m_{b}^{2} \right].
\end{equation}
Using the substitutions
\begin{equation}
u = E_{a}+E_{b} {\hskip 1 true cm \rm and \hskip 1 true cm} v = E_{a} - E_{b},
\end{equation}
the differential element changes, due to the fact that the Jacobian is
different from 1, as follows
\begin{equation}
dE_{a}dE_{b} \rightarrow {dudv  \over 2}.
\end{equation}
Completing the substitution, we get
\begin{eqnarray}
{dN^{e^{+}e^{-}}_{ab}\over d^{4}x dM^{2}} &=& {g_{ab}\over 64\pi^{4}}
\int ds\int dv \int\limits_{u_{-}}^{u_{+}} du
e^{-\beta v} {d\sigma_{ab}^{e^{+}e^{-}} \over dM^{2}}
\lambda^{1/2}(s,m_{a}^{2},m_{b}^{2}),
\end{eqnarray}
where
\begin{equation}
u_{\pm} = {v(m_{a}^{2}-m_{b}^{2}) \over s} \pm
\left({\lambda^{1/2}(s,m_{a}^{2},m_{b}^{2}) \over s}
\right) \sqrt{v^{2}-s}.
\end{equation}
After integrating over $u$, the remaining $v$ integration can only be
written in terms of a modified Bessel function.  The rate is
\begin{eqnarray}
{dN^{e^{+}e^{-}}_{ab}\over d^{4}x dM^{2}} &=& {g_{ab}\over 32\pi^{4}}
\int ds \lambda(s,m_{a}^{2},m_{b}^{2})
{{\cal K}_{1}(\beta \sqrt{s}) \over \beta \sqrt{s}}
{d\sigma_{ab}^{e^{+}e^{-}} \over dM^{2}}.
\end{eqnarray}
Introducing the dimensionless variable $z=\sqrt{s}/T$, we arrive at the
final expression
\begin{eqnarray}
{dN^{e^{+}e^{-}}_{ab}\over d^{4}x dM^{2}} &=& {T^{6}g_{ab}\over
16\pi^{4}} \int\limits_{z_{min}}^{\infty} dz {
\lambda(z^{2}T^{2},m_{a}^{2},m_{b}^{2}) \over T^{4}}
{\cal K}_{1}(z) {d\sigma_{ab}^{e^{+}e^{-}} \over
dM^{2}}(z)
\label{eq:finalgenrate}
\end{eqnarray}
where $z_{min} = (m_{a} + m_{b} + M)/T$.

\figure{Lepton pair production through virtual bremsstrahlung in
scattering of particles $a \ b \rightarrow c \ d$.  The
particles might be charged or uncharged pions, quarks or antiquarks
or gluons. The shaded region indicates a strong
interaction.\label{generice+e-}}
\figure{The interference function $f(s)$ as it changes with the
dimensionless variable $\sqrt{s}/m$.\label{plotfofs}}
\figure{(Approximate) $e^{+}e^{-}$ production rates through
virtual bremsstrahlung from pions as compared with quartic pion
interaction predictions. The dashed lines are our estimations
with varying values for the upper limit on the energy and the solid
line is from the quartic interaction.\label{chiral}}
\figure{The total elastic $\pi^{+}\pi^{-}$ cross section as
compared with experimental data from
Refs.~\cite{vs76} and ~\cite{sdp73}.\label{pipidata}}
\figure{$e^{+}e^{-}$ production rates from pions using the
exact electromagnetic factor and the model-calculated differential
cross section (solid curve) as compared to the rate using the
approximate electromagmetic expression and the assumed-symmetric
parametrization of the total cross section (dashed curve).\label{compirate}}
\figure{$t$-channel gluon exchange diagram.
\label{gluonxchng}}
\figure{Quark-driven rates at temperatures $T=200$ and 300 MeV
along with the pion driven rate at T=200 MeV using the approximate
electromagnetic
factor applicable to each process.  The long-dashed curves are
for current quarks, the dot-dashed curves are medium quark results
and the short-dashed curve (also at $T=200$ MeV) is for constituent
quarks.\label{approxqrates}}
\figure{Quark-driven rates superimposed on the pion-driven rate
but this time using the correct electromagnetic
factor. The labelling scheme is the same the previous
figure.\label{quarkrates}}
\figure{The ratio
$R\equiv |\epsilon\cdot J|^{2}_{exact}/|\epsilon\cdot J|^{2}_{approx}$
as it depends on the center-of-mass scattering angle
$\cos \theta$.\label{edjrat}}
\figure{Production rates for zero (total) momentum dileptons: The solid
curve is a parametrization of the results from Ref.~\cite{ebrpty90}, the
dashed curve is our soft-photon approximate results and the dotted
curve is the result one gets by modifying the quark dispersion
relation.\label{resumm}}
\figure{The total $e^{+}e^{-}$ yield from competing sources.
The solid curve is for the pion-driven yield,
the short-dashed curves are Dalitz decays, the long-dashed curve
is the Born approximate $q\bar q$ annihilation result and the
dot-dashed curves are I) current quarks, II) medium quarks and
III) constituent quarks.\label{totalyield}}
\figure{The total $e^{+}e^{-}$ yield with different relevant
temperatures.\label{totalyield2}}
\figure{The total $e^{+}e^{-}$ yield with still a different set
of relevant
temperatures.\label{totalyield3}}
\end{document}